\documentclass[final,1p,times]{elsarticle}

\usepackage{amsmath,amssymb,amsthm,mathtools}
\usepackage{bm}
\usepackage{graphicx}
\usepackage{hyperref}
\usepackage{booktabs}
\usepackage{physics}
\usepackage{siunitx}
\usepackage{color}
\usepackage[utf8]{inputenc}
\usepackage[T1]{fontenc}
\usepackage[portuguese,english]{babel}
\usepackage{amssymb}
\usepackage{amsmath}
\usepackage{amsthm}
\usepackage{xcolor}
\usepackage{natbib}
\usepackage{hyperref}
\usepackage{subcaption}
\usepackage[utf8]{inputenc}
\usepackage[T1]{fontenc}
\usepackage{amssymb}
\usepackage{amsmath}
\usepackage{amsthm}
\usepackage{xcolor}
\usepackage{natbib}
\usepackage{hyperref}
\usepackage{subcaption}

\journal{Annals of Physics}

\begin{document}

\begin{frontmatter}



\title{Emergent $\Lambda$CDM cosmology from a measure-induced deformation of the Newtonian action}

\author[1]{S. M. M. Rasouli}
\ead{mrasouli@ubi.pt}
 \address[1]{Departamento de F\'{i}sica,
Centro de Matem\'{a}tica e Aplica\c{c}\~{o}es (CMA-UBI),
Universidade da Beira Interior,
Rua Marqu\^{e}s d'Avila
e Bolama, 6200-001 Covilh\~{a}, Portugal}

\begin{abstract}
We propose a minimal extension of the Newtonian action by introducing a time-dependent fractional kernel characterized by a single deformation parameter $\alpha$. This kernel admits a natural interpretation as a nontrivial integration measure defined by a time-dependent kernel, placing the formulation within measure-based approaches to anomalous or fractal dynamics. Despite the appearance of a friction-like term in the equations of motion, a conserved quantity is still obtained, containing a memory-like fractional kinetic energy contribution. 
Moreover, by generalizing the standard Newtonian potential to an effective $\alpha$-dependent potential induced by the underlying measure, the resulting cosmological equations exhibit an effective correspondence with relativistic FLRW cosmology at the level of background dynamics. In the limit $\alpha=1$, the framework reduces to standard Newtonian cosmology. We show that, with a single unified potential, the matter-dominated, radiation-dominated, and present accelerated phases are obtained self-consistently, while the latter two epochs cannot be described within standard Newtonian cosmology. 
The structural presence of $\alpha$ in all physical observables allows theoretical and observational constraints to be imposed, indicating compatibility with observational data in the regime where $\alpha$ is close to unity. 
Within this framework, an effective cosmological constant naturally arises, controlled by the small deviation of $\alpha$ from the Newtonian limit. These results show that the proposed fractional framework can effectively reproduce the main background dynamical features of $\Lambda$CDM cosmology through a simple measure-induced deformation of the Newtonian action.

\end{abstract}

\begin{keyword}
Modified gravity \sep Fractional dynamics \sep Measure-based dynamics \sep Lebesgue-Stieltjes integral \sep Dark energy \sep Relativistic and Newtonian cosmologies \sep $\Lambda$CDM model
\end{keyword}

\end{frontmatter}


\section{Introduction}
\label{Intro}

The Newtonian laws are not only remarkably powerful in describing the majority of phenomena around us, but are also mathematically elegant and simple in form. Isaac Newton's  great achievement was to unify these insights into a coherent mathematical framework of universal applicability \cite{newton1987philosophiae}. Almost a century later, these laws were reformulated by Joseph-Louis Lagrange and William Rowan Hamilton, within the more abstract framework of the principle of least action, a simple but profound philosophical idea \cite{yourgrau2012variational,safko2002classical,landau2013mechanics}. This variational reformulation provided not only mathematical elegance but also a powerful tool for generalization to modern physics.

Despite the astonishing simplicity, elegance, and predictive power of Newton's laws, they nevertheless fail to describe certain important phenomena. Beyond their empirical success, these laws were also grounded in philosophical assumptions, such as the concepts of absolute space and absolute time,  which were questioned by philosophers such as Leibniz and later by Mach \cite{d2022introducing}. The recognition of these shortcomings, both empirical and conceptual, paved the way for the birth of what we now call modern physics: special and general relativity, and quantum mechanics.

In these modern frameworks, physicists did not discard Newtonian mechanics as wrong, but rather reinterpreted it as a limiting case valid within a restricted domain (Newtonian domain), namely, for moderate velocities, weak gravitational fields, and macroscopic scales. This point of view led to the \textit{correspondence principle}, according to which any prediction of a modified fundamental theory might be reduced to that of the corresponding Newtonian mechanics under the appropriate conditions \cite{kragh2012niels,landau2013quantum}. 

Moreover, there are boundary cases that highlight the shortcomings of Newton's dynamics. The most famous example is the anomalous precession of Mercury's perihelion. Although Newtonian mechanics, together with planetary perturbations, can explain most of the orbital motion of the planets, it consistently failed to account for a small but measurable excess precession in Mercury's orbit. This puzzle, lying precisely at the border between Newtonian mechanics and relativity, was fully explained by Einstein's general relativity (GR) \cite{landau1963classical,thorne2000gravitation}.

In what follows, we argue that certain conceptual questions and observations can independently motivate a reconsideration and extension of Newtonian dynamics. Importantly, such an extension cannot be considered as an alternative to modern theories but rather serves as a phenomenological generalization within the classical Newtonian framework.

\begin{enumerate}

\item 

In real-world systems, not only conservative forces but also non-conservative forces play a crucial role in many physical phenomena. However, within standard Newtonian mechanics, only conservative forces admit variational formulation based on the standard action. Non-conservative forces, such as friction and drag, are typically introduced directly at the level of the equations of motion \cite{landau1987fluid,white2006viscous}.
This raises the question of whether the classical action can be generalized to consistently incorporate both types of forces within a unified variational framework 
\cite{lanczos2012variational,landau2013mechanics}.

\item 

One may further ask whether Newton's second law can be 
generalized through derivatives of non-integer order, while still reproducing known phenomena or extending their range of validity \cite{Laskin:2000sed,dong2008space,Rasouli:2021lgy}? In a similar spirit, it is natural to consider whether the Newtonian gravitational potential can be generalized to describe effects beyond its standard domain of applicability.

\item 

The principle of least action provides a universal and formal foundation for classical mechanics. In its standard form, the action is local and Markovian, leading directly to energy conservation through time-translation invariance. This naturally prompts the question of whether a modified action can retain the variational structure while allowing for nonlocal or dissipative effects and still admit a well-defined conserved quantity.

\item 
Whether Newtonian mechanics can be generalized in such a way as to remain valid in its usual domain while also providing, at least phenomenologically, some description of phenomena lying slightly outside it. For example, Newtonian cosmology (NC), first developed in the early 20th century by E.A. Milne and W.H. McCrea  \cite{milne1934newtonian,mccrea1934newtonian,mccrea1955newtonian}, successfully reproduces the standard Friedmann equations in the particular case under certain assumptions. Could such a framework be further generalized to describe the early time inflation, radiation-dominated epoch, and the present accelerated expansion of the universe? 
\end{enumerate}

The nature of these questions easily leads the inquisitive mind to 
reformulate Newton's equations in 
a way that overcomes some of its most obvious shortcomings while maintaining their success in the Newtonian domain.

In this work, inspired by fractal spacetime frameworks, where the standard integration measure is replaced by a nontrivial Stieltjes measure \cite{calcagni2010quantum,calcagni2010fractal}, we consider a corresponding deformation at the level of classical dynamics. 
Concretely, we generalize the Newtonian action by incorporating a time-dependent kernel and an effective potential. This effective model yields modified equations of motion that reduce to the standard Newtonian form in the limit $\alpha=1$, where $\alpha$ is the fractional parameter.

It is worth noting that the use of fractional calculus to generalize standard quantum mechanics has already been investigated (see, for example, \cite{laskin2000fractional,muslih2010fractional,iomin2011fractional,Rasouli:2021lgy,lim2021editorial,duan2024uniform,varao2024fractional,umar2025tunneling}, and references therein). Fractional modifications of classical and quantum field theories and gravitational dynamics have been systematically investigated \cite{calcagni2021classical,calcagni2021quantum}.
Moreover, fractional frameworks have also been applied to relativistic cosmology, both by the present author (see, e.g., \cite{Rasouli:2022bug,deOliveiraCosta:2023srx,rasouli2024fractional}) and by other researchers \cite{landim2021fractional,el2024schwarzschild,garcia2022cosmology,socorro2023quantum,ccoker2023modified, micolta2024fractional,el2024fractional,micolta2025fractional,jalalzadeh2025fractional}. 

The main motivation for introducing the modified classical action considered in this work is the investigation of gravitational and cosmological dynamics at large scales. In this spirit, we seek an effective extension of NC that allows us to revisit, at least from a phenomenological perspective, a broad class of problems studied within modern cosmology and various extended cosmological frameworks, including early universe scenarios, late-time cosmological dynamics and dark energy phenomenology~\cite{doroud2009class,rasouli2016exact,rasouli2020late,Rasouli:2022hnp,ildes2023analytic,rasouli2024phase,khan2026long,khan2026dilaton,khan2026post}. The proposed formulation is therefore designed to provide a controlled and exploratory generalization of Newtonian cosmology (NC), aimed at probing its scope and limitations in regimes relevant to large-scale gravitational dynamics.

The structure of this paper is organized as follows. 
In the next section, we briefly review and categorize the main approaches that have been proposed for generalizing Newton's laws using fractional procedures. 
We then introduce a modified classical action incorporating a time kernel and an effective potential, from which the fractional Newtonian equations of motion are derived and the resulting dynamics is analyzed. In Section \ref{Frac-Cosmology},  we apply the present fractional framework to cosmology and show that the main equations governing relativistic background cosmology can be effectively reproduced within this approach.
In Section \ref{LCDM}, employing the fractional Newtonian cosmological framework, we
recover exact background solutions corresponding to the $\Lambda$CDM model
across different cosmic epochs. In addition, we explicitly derive the various
quantities of the model, demonstrating that our cosmological framework is fully
self-consistent, both at the level of the governing equations and at the level
of the resulting solutions.
In Section \ref{unif-pot}, motivated by the results obtained in Section \ref{LCDM}, we propose a unified
effective potential. We show that within this framework, the
$\Lambda$CDM-like background solutions can be recovered consistently. Moreover, the
proposed potential allows the fractional effects to be clearly
isolated, thereby enabling meaningful constraints to be placed on the
fractional sector using recent cosmological observations.
Finally, in Section \ref{Concl}, we summarize our
main results and provide a further discussion on the 
physical implications of the fractional NC under investigation.

\section{Fractional extension of standard Newtonian action and equations of motion}
\label{Frac-New-Mech}

In order to extend the classical Newtonian mechanics into a fractional framework, we may apply three main approaches:
\begin{itemize}
    \item 

 The classical derivative appearing in the classical action is directly replaced with a specific fractional derivative, such as the Riemann–Liouville or Caputo derivative \cite{riewe1996nonconservative,riewe1997mechanics}.

\item  The time derivatives in standard Newtonian equations are 
substituted with fractional ones. Although this method is mathematically less rigorous from a variational perspective, it has proven effective in modeling anomalous phenomena in a wide range of physical systems \cite{chung2015fractional,varieschi2017applications,elzahar2020generalized}.

\item A third approach consists in modifying the classical action by introducing a time-weighted kernel. In this formulation, fractional effects are incorporated directly at the level of the action, without the need to introduce fractional derivatives explicitly. Importantly, such a kernel can be naturally interpreted as defining a nontrivial integration measure in time, namely $d\varrho(\tau)=v(\tau)\,d\tau$ with $v(\tau)\propto(\bar t-\tau)^{\alpha - 1}$, where $\alpha>0$ is a fractional parameter. This perspective is consistent with measure-based approaches to anomalous or fractal dynamics \cite{calcagni2010quantum} (see also \cite{calcagni2010fractal}), where modifications of the integration measure effectively encode nonstandard scaling properties of the underlying system. In this sense, the present construction can be viewed as a minimal realization of a nontrivial temporal geometry, while preserving the standard variational structure of classical mechanics.

This approach has been applied across a wide range of investigations including relativistic cosmology; see, for instance, \cite{shchigolev2013fractional,shchigolev2016testing,el2017fractional,micolta2023revisiting,gonzalez2023exact,rasouli2024fractional} and references therein.

\end{itemize}

 In this paper, we focus exclusively on the third approach, as it allows for a conceptually clean modification of Newtonian dynamics via an extended variational principle, avoiding the mathematical complexities of nonlocal derivatives while still capturing key memory effects and non-conservative features of fractional dynamics.

 Let us start with a generalized form of the classical action functional for a single non-relativistic particle of mass $m$ moving in three spatial dimensions.
 Concretely, we consider a minimal deformation of Newtonian dynamics in which the action is weighted by a fractional time-dependent kernel governed by a single parameter $\alpha$:

\begin{eqnarray}\label{New-fr-action}
 S_{\alpha}=
 \!\!\frac{1}{\Gamma(\alpha)}\int_{0}^{\bar t} v({\tau} )\left(T-V_{\rm eff}\right)\,d{\tau},
\end{eqnarray}
where 
\begin{equation}\label{L-xi}
 v({\tau})\equiv\left(\frac{{\bar t} - \tau}{\tau_0}
\right)^{\alpha-1}.
\end{equation}
In equations \eqref{New-fr-action} and \eqref{L-xi}, $\tau_0$ is a fixed reference time scale introduced solely to render the kernel $ v$ dimensionless; the vector $\mathbf{r}(\tau)$ represents the trajectory of the particle; $T=\frac{1}{2}\,m (\frac{d {\mathbf{r}}}{d\tau} )^2$ is the standard kinetic energy; $V_{\rm eff}=V_{\rm eff}\left(\mathbf{r};\alpha\right)$ represents an effective potential energy that, in the limit $\alpha=1$, consistently reduces to the corresponding standard (Newtonian) potential energy $V(\mathbf{r})$ associated with conservative forces.
Therefore, in a particular case where $\alpha=1$, the action \eqref{New-fr-action} reduces to the usual classical action of Newtonian mechanics. 

The form of the action \eqref{New-fr-action}
can be interpreted as a Stieltjes-type integral \cite{calcagni2010quantum,calcagni2010fractal,carter2000lebesgue}, where the weight function $v(t)$ defines a nontrivial temporal measure $d\varrho(\tau)=v(\tau)\,d\tau$. A more detailed justification of this choice of action will be presented.

From the action \eqref{New-fr-action}, we can
easily obtain the equations of motion as
\begin{eqnarray}
m\,\ddot{\mathbf{r}}
= -\nabla V_{\rm eff}(\mathbf{r})- m \gamma_\alpha(t) \,\dot{\mathbf{r}}, \qquad \gamma_\alpha(t)\equiv\frac{(\alpha - 1)}{t},
\label{fr-eq-3d}
\end{eqnarray}
where we used the transformation $\bar{t}-\tau\equiv t$,  and a dot denotes the derivative with respect to $t$; $\gamma_\alpha(t)\equiv \dot{v}/{v}$, and $\nabla$ is the gradient operator. 
 We should note that, throughout this work, any quantity that carries the subscript $\alpha$ explicitly depends on the fractional parameter $\alpha$.

At this point, it is important to stress that the choice of the action \eqref{New-fr-action} is not ad hoc, but is motivated by measure-based formulations of spacetime geometry in gravitational models.
 The idea of modifying the underlying geometric structure of spacetime through a nontrivial integration measure has been extensively developed \cite{calcagni2010quantum,calcagni2010fractal}, where the standard Lebesgue measure $d^Dx$ is replaced by a generalized Stieltjes-type measure $d\varrho(x)=v(x)d^Dx$. In this framework, the function $v(x)$ encodes a scale-dependent (fractal) structure of spacetime, leading to an effective reduction of the Hausdorff dimension at short distances and improved ultraviolet behavior. A key consequence of this construction is the appearance of additional terms of the form $\partial_\mu v / v$ in the equations of motion, which effectively act as friction-like contributions while having a purely geometric origin.

In the present work, we consider a simplified realization of this idea within a Newtonian framework by introducing a time-weighted action functional. 
In particular, the fractional kernel $\left[({\bar t} - \tau)/{\tau_0}\right]^{\alpha-1}$ can be interpreted as a Stieltjes-type measure in time, so that the action takes the form \eqref{New-fr-action} with \eqref{L-xi}. This construction can be viewed as an effective one-dimensional (temporal) reduction of Calcagni's measure-based framework, in which the deformation is restricted to the temporal integration measure. As a result, the equations of motion acquire an additional term proportional to $ m \gamma_\alpha(t) \,\dot{\mathbf{r}}$, which mirrors the structure of the measure-induced contributions in the full field-theoretic setting.

Furthermore, the dependence of the effective potential energy $V_{\rm eff}=V_{\rm eff}\left(\mathbf{r};\alpha\right)$  on the parameter $\alpha$ can be naturally justified within this perspective. Since the weight function $v(t)$ and thus the effective measure explicitly depend on $\alpha$, the underlying geometric background becomes $\alpha$-dependent. Consequently, physical quantities derived from the action, including interaction terms and energy contributions, are expected to inherit this dependence. In this sense, the $\alpha$-dependent potential should be interpreted not as an ad hoc modification, but as an effective description of how the fractal-like temporal geometry influences the dynamics of the system.

As seen in fractional Newtonian equations, in addition to the potential force $\mathbf{F}_{\rm pot}\equiv-\nabla V_{\rm eff}(\mathbf{r};\alpha)$, an additional drag-like term appears of the form $\mathbf{F}_{\alpha}^{\rm drag} \equiv m \gamma_\alpha(t) \,\dot{\mathbf{r}}$. This term can be interpreted as a dissipative force that is linear in velocity and explicitly time-dependent, suggesting a friction- (anti-friction-) like behavior. Moreover, other force terms receive modifications encoded in the effective potential.

In classical physics, linear drag forces are well known and are typically modeled as $\mathbf{F}_{\text{drag}} = F_0 \, \dot{\mathbf{r}}(t)$, where $F_0$ is a constant.
Such forces appear in problems involving motion through viscous media, such as falling spheres in air or motion in fluids. However, these forces cannot be derived from a standard action principle, due to their non-conservative nature. In fact, Rayleigh was among the first to manually include such dissipative forces in the classical Lagrangian dynamics \cite{riewe1996nonconservative,safko2002classical}.

Dissipative forces in nature often exhibit more complex 
structures, being non-linear functions of velocity or even depending explicitly on time. For example, the quadratic drag force 
is used in modeling projectiles moving through the atmosphere. In more complex systems, such as non-Newtonian fluids, viscoelastic media, or biological systems with memory, frictional forces may depend on both velocity and time in a highly non-linear fashion.

In light of the above, one of the most notable features of our model is that a time- and velocity-dependent dissipative force naturally emerges from a generalized fractional action. Even more importantly, a memory-dependent effective potential can still be defined, leading to a generalized mechanical energy, which remains conserved in the absence of external forces. As we show below, this fractional mechanical energy contains an additional term representing the logarithmic time average of the kinetic energy, which can be interpreted as a fractional kinetic energy. Unlike Rayleigh's model, our approach may respect the correspondence principle: in the limit $\alpha = 1$, the dissipative term vanishes, and the model exactly reduces to standard Newtonian dynamics. Moreover, we emphasized that although the standard Rayleigh-type dissipative forces cannot be derived from a variational principle in classical mechanics, in our framework such forces naturally emerge from the fractional action. 
These include time-dependent drag terms that resemble those discussed in \cite{riewe1996nonconservative}. Not only does this provide a theoretically grounded framework for modeling dissipative systems, but it also opens up new directions in fractional NC, as we will elaborate in the following.

As follows from the action \eqref{New-fr-action} and the corresponding equations of motion \eqref{fr-eq-3d}, in our fractional model, the energy exchanges between the standard and fractional sectors are such that
\begin{eqnarray}
\frac{dE^{^{\rm mech}}_{\rm st}}{dt}=\mathcal{F}(\alpha)\neq0, \hspace{5mm} E^{^{\rm mech}}_{\rm st}\equiv T+V(\mathbf{r}),
\label{con-eq}
\end{eqnarray}
which indicates that the standard 
mechanical energy, $E^{^{\rm mech}}_{\rm st}$, is generally not conserved. 
As explicitly shown, the presence of a time dependent kernel leads to the breakdown of the conservation of the standard mechanical energy. Similar situations are well known in classical dissipative systems, such 
as those described by the Caldirola-Kanai action \cite{um2002quantum,kim2003squeezed, vestal2021bateman}, where the standard mechanical energy is not conserved, yet a 
consistent variational formulation can still be constructed.

As clearly observed, the main reason for the deviation from the standard Newtonian model, even by assuming $V_{\rm eff}=V$, lies in the presence of a time-dependent damping term in the equations of motion, which explicitly breaks the conservation of mechanical energy. More concretely, the fractional action \eqref{New-fr-action} is not invariant under time translations. We should note that this feature is not unique to our fractional framework herein. For instance, in a gravitational model within the context of general relativity, where a scalar field is minimally coupled to gravity, the introduction of a time-dependent kernel also leads to a violation of energy-momentum conservation for the scalar field \cite{rasouli2024fractional,Rasouli:2025qix}. This behavior can be interpreted as an exchange of energy between the scalar field and the underlying geometry or effective background (i.e. the fractional sector), where we have shown that even in such scenarios, the second Noether theorem still holds in a geometrical sense, both at the background level and at the level of first-order perturbations \cite{Rasouli:2025qix}. In our fractional model, as we shall show in the following, it is still possible to obtain a conserved quantity.

In the present framework, instead of abandoning the concept of mechanical energy conservation altogether, we seek a quasi-standard conserved quantity that allows for a meaningful redefinition of an effective mechanical energy.
This requirement naturally motivates the introduction of an additional potential term in the action.
Consistency also demands that this potential share the same fractional parameter $\alpha$ that controls the time-dependent kernel, so that the entire fractional sector switches off smoothly in the limit $\alpha = 1$, ensuring the full recovery of the corresponding standard model.

By taking the dot product of both sides of \eqref{fr-eq-3d} with $\dot{\mathbf {r}}$ and performing a straightforward computation, we obtain
\begin{eqnarray}\label{Fr-cons}
\mathcal{E^{^{\rm mech}}_{\rm eff}}\equiv T + V_{\rm eff} +T_\alpha = \text{constant},
\end{eqnarray}
where 
\begin{eqnarray}\label{frac-T}
T_\alpha \equiv  \, 2( \alpha-1) \int_{t_i}^{t_f} T(t)\, d\ln t,
\end{eqnarray}
and $\mathcal{E^{^{\rm mech}}_{\rm eff}}$ can be interpreted as the \textit{effective (total) mechanical energy}, which remains conserved in the absence of external forces. 
Although the explicit time dependence of the measure modifies the standard notion of time-translation invariance, a generalized conserved quantity still emerges from the variational structure.

In relation \eqref{Fr-cons}, we have intentionally introduced the symbol $T_\alpha$, which admits an interesting interpretation: This quantity may be interpreted as a logarithmic-time average of the kinetic energy between the initial and final times, rescaled by the fractional deviation $(\alpha-1) $. 
Such a structure bears a close resemblance to the logarithmic memory effects that emerge in certain non-local or history-dependent scenarios in classical and quantum dynamics.  
In the standard limit, where $\alpha = 1 $, the fractional sector of $V_{\rm eff}$ and  $T_\alpha$ disappear, and the standard Newtonian conservation law is recovered, see equation \eqref{Fr-cons}.

In a specific case,  the effective potential can be decomposed as the sum of the standard part and the fractional correction, $V_{\rm eff}=V_{\rm st}+V_{\alpha}$, for which equation \eqref{Fr-cons} becomes
\begin{eqnarray}\label{Fr-cons-1}
\mathcal{E^{^{\rm mech}}_{\rm eff}}=E^{^{\rm mech}}_{\rm st}+E^{^{\rm mech}}_{\alpha}
=(T+ V )
+(T_\alpha+V_{\rm \alpha}) = \text{constant}.
\end{eqnarray}
The structural motivation for introducing the additional potential $V_\alpha$ into the action now becomes transparent. The fractional quantities $T_\alpha$ and $V_{\rm \alpha}$ together encode the fractional contribution to the dynamics,  playing the role of the fractional mechanical energy sector that modifies the standard one. The rationale for including an explicit fractional potential term $V_{\rm \alpha}$ is twofold. First, the time kernel deforms the dynamical sector and generically induces nonstandard (fractional) kinetic contributions, implying that the conventional mechanical energy is not conserved in its standard form. Second, in order to maintain a consistent and phenomenologically viable weak-field central-force description (while preserving spherical symmetry), one is naturally led to allow for a $\alpha$-dependent correction in the interaction sector as well. Importantly, this does \emph{not} introduce new dynamical degrees of freedom: the configuration variables remain $(r,\phi)$ (or equivalently, the planar Cartesian coordinates), and $\alpha$ is a fixed deformation parameter rather than a dynamical field. Thus, the model remains a two degree of freedom central-force system with a single controlling parameter $\alpha$, admitting the exact Newtonian limit as $\alpha\to1$.

  In summary, by extending the standard Newtonian action through the introduction of a time-weighting kernel depending on the fractional parameter $\alpha$, an effective fractional sector naturally emerges in the dynamics. This sector may be interpreted as a memory-like kinetic contribution that fundamentally modifies the temporal structure of the theory. 
  In addition, inspired by the same measure, to preserve internal consistency, self-coherence, and structural elegance of the framework, a corrective potential $V_\alpha$ is introduced as a fractional modification of the gravitational interaction. Importantly, no new dynamical degrees of freedom are added: the dynamical variable remains $\mathbf r(t)$, while the parameter $\alpha$ acts purely as a structural deformation parameter controlling deviations from the Newtonian limit.

In the next section, we show that the resulting cosmological equations exhibit a structural correspondence with those of relativistic cosmology, namely the equations derived from the Einstein-Hilbert action for the FLRW metric.

\section{Structural correspondence between fractional Newtonian and relativistic cosmological equations}
\label{Frac-Cosmology}

Even after the advent of general relativity and the development of relativistic cosmology, research on Newtonian cosmology (NC) has continued to attract considerable attention \cite{milne1934newtonian,mccrea1934newtonian,mccrea1955newtonian,mccrea1955significance,callan1965cosmology,jordan2005cosmology,ellis2013discrete,tipler1996newtonian,Gouba:2021bjs}. This sustained interest is mainly motivated by the mathematical simplicity of Newtonian dynamics, the relative transparency of its physical assumptions, and the structural resemblance of the resulting equations to those obtained in relativistic cosmology at both background and perturbative levels, especially in the matter-dominated regime within specific limits~\cite{mccrea1951relativity, green2012newtonian}.
Nevertheless, this resemblance should be regarded as an analogy rather than a strict equivalence. NC is fundamentally constructed upon absolute time and a flat Euclidean space, which are conceptually distinct from the relativistic spacetime framework.

As a consequence, NC suffers from several well-known limitations. These include the assumption of a globally Euclidean spatial geometry (for a Newtonian cosmology on a non-Euclidean absolute space, see, for instance, \cite{Barrow:2020wmk,vigneron2022non}),  the absence of horizons and causal structure,  the inability to consistently describe accelerated expansion, and ambiguities in defining corresponding conserved quantities such as energy and momentum in cosmological context.
 For these reasons, the standard NC cannot be viewed as a genuine alternative to relativistic cosmological models. Instead, it should be interpreted as a phenomenological and educational framework that offers valuable insights in the classical, non-relativistic regime of cosmological dynamics. Nevertheless, it should be noted that Ref. \cite{Faraoni:2020uuf} provides an insightful perspective on previously overlooked analogies between standard NC and the corresponding relativistic setting.

In this work, employing the framework established in the previous section, we propose a fractional extension of Newtonian cosmology. 
The main objective is to investigate whether suitably modified fractional dynamics can consistently reproduce the key structural and dynamical features of the relativistic cosmological evolution.

To this end, we consider a homogeneous and isotropic spherical region of physical radius $a(t)$, filled with a pressure-less fluid 
(i.e., non-relativistic matter) with density $\rho_m(t)$. 
Within this region, the Newtonian gravitational potential $\Phi_N$ 
satisfies the Poisson equation, which for a uniform matter distribution leads to
\begin{equation}\label{S-Mass}
    \nabla_{\!r}\,\Phi_{\rm N}=\frac{4 \pi G}{3} \rho_m \mathbf{r},
\end{equation} 
where $\nabla_{\mathbf r}$ denotes the gradient with respect to the physical coordinates and $\mathbf r$ is the position vector measured from the center of the sphere. 

A test particle with mass $m$ is assumed to be located on the surface of the sphere, and its dynamics will be used to derive the cosmological equations of the model.  Using the fractional equation \eqref{fr-eq-3d}, we obtain 
\begin{eqnarray}\label{acc-eq}
 \ddot{a}(t)+(\alpha-1)\frac{\dot{a}(t)}{t}+\frac{d\Phi_{\rm eff}(a; \alpha)}{da}=0,
\end{eqnarray}
where the second term encodes memory effects induced by the time dependent kernel, $\Phi_{\rm eff}(a;\alpha)$ denotes the effective gravitational potential and we assumed $m=1$ for simplicity. 

In this work, for the sake of simplicity, we assume that the effective potential is composed of the sum of Newtonian gravitational potential $\Phi_{_{N}}$ and the fractional potential $\Phi_{\alpha}$ as
\begin{eqnarray}\label{eff-pot-cos-0}
\Phi_{\rm eff}=\Phi_{_{\rm N}}+\Phi_{\alpha},
\end{eqnarray}
where $\Phi_{\alpha}$ is assumed to vanish in the limit $\alpha=1$, ensuring recovery of standard NC.

To construct our fractional cosmological model, let us also define the following quantities
\begin{eqnarray}\label{cos-T-def}
T&\equiv& \frac{1}{2}\dot{a}^2, \hspace{10mm}
T_{\alpha}\equiv
2(\alpha-1)\int_{t_i}^{t_f}T(t)\, d\ln t,\\\nonumber\\
\label{cos-E-def}
E&\equiv& T+\Phi_{_{N}}, \hspace{5mm} E_{\alpha}\equiv T_{\alpha}+\Phi_{\alpha},
\end{eqnarray}
where, for the particular case $\alpha=1$, the quantity $E_{\alpha}$ vanishes. 

Using equations \eqref{Fr-cons}, \eqref{S-Mass}, \eqref{eff-pot-cos-0} and \eqref{cos-T-def}, and \eqref{cos-E-def}, we obtain the fractional Friedmann equation:
\begin{eqnarray}\label{eff-frd}
 H^2=\frac{8 \pi G}{3}\rho_{_{\rm eff}}-\frac{\mathcal{K}}{a^2},
\end{eqnarray}
where
\begin{eqnarray}\label{K-def}
\mathcal{K}&\equiv& -2 \mathcal{E}=-2(E+E_{\alpha})=\mathrm{constant}, \\
\label{rho-eff}
\rho_{_{\rm eff}}&\equiv&\rho_m+\rho_{\alpha}, \hspace{5mm}\rho_{\alpha}\equiv-\frac{3}{4 \pi G}\left(\frac{E_{\alpha}}{a^2}\right).
\end{eqnarray}
Several important points regarding equations \eqref{eff-frd}, \eqref{K-def} and \eqref{rho-eff} should be highlighted: (i) $\mathcal{K}$ should not be interpreted as a spatial curvature parameter, but rather as an integration constant associated with the total effective energy of the system; (ii) although equation  \eqref{eff-frd} is formally resembles the relativistic Friedmann equation, it is entirely derived within the fractional Newtonian framework; (iii) $\rho_{\alpha}$ should be interpreted as an effective energy density induced by the fractional sector, rather than a physical ordinary matter component.

Using equation \eqref{acc-eq} and the definitions \eqref{cos-T-def}, \eqref{cos-E-def} and \eqref{rho-eff}, we obtain the fractional acceleration equation:
\begin{eqnarray}\label{acc-eff-eq}
\frac{ \ddot{a}}{a}=-\frac{4\pi\,G}{3}
\Big(\rho_{_{\rm eff}}+3 \,p_{_{\rm eff}}\Big),
\end{eqnarray}
where  
\begin{eqnarray}\label{p-eff}
p_{_{\rm eff}}=p_{\alpha}\equiv\frac{1}{4\pi\,G}\frac{1}{a^2}
\frac{d}{da}\Big(a\,E_{\alpha}\Big),
\end{eqnarray}
denotes the effective pressure, which emerges purely from the fractional sector and does not correspond to microscopic interactions.

Employing equations \eqref{eff-frd} and \eqref{acc-eff-eq}, one can easily obtain a conservation law as
\begin{eqnarray}\label{cons-eq-cos}
\dot{\rho}_{_{\rm eff}}+3H (\rho_{_{\rm eff}}+p_{_{\rm eff}})=0.
\end{eqnarray}

Moreover, using the relations \eqref {rho-eff} and \eqref{p-eff}  one 
can easily show that the fractional energy density  $\rho_\alpha$ 
and pressure $p_{\alpha}$ obey the following conservation equation:
\begin{eqnarray}\label{cons-eq-fr}
\dot{\rho}_{\alpha}+3H (\rho_{\alpha}+p_{\alpha})=0.
\end{eqnarray}
Furthermore, using equations \eqref {cons-eq-cos} and \eqref{cons-eq-fr}, we obtain the 
same continuity equation of the standard NC:
\begin{eqnarray}\label{cons-eq-rho-m}
\dot{\rho}_{\rm m}+3H\rho_{\rm m}=0.
\end{eqnarray}

Let us define an effective equation of state (EoS) parameter as
\begin{eqnarray}\label{w-def}
w_{_{\rm eff}}\equiv\frac{p_{_{\rm eff}}}{\rho_{_{\rm eff}}}=(4 \pi G a^2 \rho_m-3\, E_\alpha)^{-1}\,\frac{d(a\, E_{\alpha})}{da},
\end{eqnarray}
where we used relations \eqref{rho-eff} and \eqref{p-eff}.

 Using the continuity equation \eqref{cons-eq-cos}, we obtain 
\begin{eqnarray}\label{w-gen}
\rho_{_{\rm eff}}(a,\alpha)=\rho_i (\alpha)\exp{\left[-3\int_{a_i}^{a}(1+w_{_{\rm eff}}(\bar{a},\alpha))\frac{d\bar{a}}{\bar{a}}\right]}, 
\end{eqnarray}
where $\rho_i$ is an integration constant and $w_{_{\rm eff}}(\bar{a},\alpha)$ is given by \eqref{w-def}.

It is important to note that equations \eqref{eff-frd}, \eqref{acc-eff-eq} and \eqref{cons-eq-cos} together form a closed dynamical system, fully determining the background evolution in our fractional NC. 

As seen, $\rho_\alpha$ and the effective pressure ${p}_{_{\rm eff}}$ fully emerge from the fractional sector, such that, in the limit $\alpha=1$, they vanish; moreover, in this limit, $\mathcal{K}\to K\equiv-2E= -2\left[T+\Phi_{_{N}}(t)\right]=\mathrm{constant}$, and therefore equations \eqref{eff-frd}, \eqref{acc-eff-eq} and \eqref{cons-eq-cos} reduce to the corresponding standard Newtonian framework.

In the standard Newtonian model, defining the escape velocity from the surface of the uniform expanding sphere as $v_\mathrm{escape}=\sqrt{\frac{2\,G \,M}{a}}$ (where $M$ is the mass of the sphere), and using equations \eqref{eff-frd}, we easily obtain three distinct dynamical regimes \cite{mccrea1934newtonian,Faraoni:2020uuf}: 
   (i) If $v>v_\mathrm{escape}$, then $E>0$ which corresponds to $K<0$. 
    (ii) If $v=v_\mathrm{escape}$, then $E=0$ which corresponds to $K=0$. 
    (iii) If $v<v_\mathrm{escape}$, then $E<0$ which corresponds to $K>0$.
Even with an appropriate choice of the size of the comoving sphere, we may recover the values $K=-1,0,1$ \cite{green2012newtonian}.
These regimes have been considered as correspondences between the standard NC and relativistic cosmology.

Let us briefly summarize the main achievements of this section. We have shown
that the resulting fractional cosmological equations coincide with the standard relativistic FLRW equations at the level of homogeneous and isotropic background dynamics.

A key and noteworthy feature of the present framework is
that it provides a bridge between standard NC and relativistic cosmology: in the special case $\alpha = 1$, the model reduces
exactly to the standard Newtonian cosmological description, while in an
appropriate complementary limit it reproduces all the dynamical signatures of
relativistic $\Lambda$CDM cosmology. Within this construction, no quantity is
introduced by hand; instead, all relations emerge self-consistently from the
underlying theoretical structure of the model.

In the next two sections, we employ the cosmological
framework proposed here and demonstrate how the background solutions
associated with different cosmic epochs can be obtained using two independent
and conceptually distinct approaches. We then analyze and compare the results
obtained from these methods in order to assess the robustness and consistency
of the framework.
We will show that both the time-dependent kernel
introduced and the fractional potential constitute the two
fundamental pillars of the framework. 

\section{Exact background solutions and reconstruction of the $\Lambda$CDM epochs}
\label{LCDM}

Exact background solutions of the relativistic cosmological field equations, both within the standard cosmological framework and in modified theories of gravity \cite{rasouli2016exact,rasouli2019kinetic,rasouli2021geodesic,rasouli2022noncommutativity,rasouli2022geodesic,ildes2023analytic}, in four and higher dimensions \cite{doroud2009class,reyes2018emergence,mousavi2022cosmological,rasouli2022noncompactified,sania2023cosmic}, have predominantly been derived using analytical or numerical approaches and serve as the fundamental basis for exploring the various phases of cosmic evolution.

Motivated by the central role played by background solutions, the main objective of this section is to show that the proposed fractional
cosmological model is capable of reproducing those
associated with different cosmic epochs in full analogy with the
$\Lambda$CDM model. In addition, we explicitly derive all relevant fractional
quantities and verify the consistency of the fundamental equations of the model
with solutions. In particular, we investigate whether the resulting background evolution can consistently reproduce the radiation-dominated, matter-dominated, and late-time accelerating phases known from relativistic cosmology. 


Before proceeding, we emphasize that the present framework is intended as an effective description at the level of background cosmological dynamics, rather than a fundamental relativistic theory.

From now on, we restrict our analysis to the case $\mathcal{K}=0$, which allows for a direct comparison with spatially flat relativistic cosmology and is also well supported by current cosmological observations \cite{Planck:2018vyg}. 

Before entering the following subsections, we briefly discuss the role played by the ordinary matter in our model. Using the standard continuity equation \eqref{cons-eq-rho-m}, we obtain:
\begin{eqnarray}\label{dens-m}
\rho_{\rm m}(a)= \rho_i\left(\frac{a}{a_i}\right)^{-3}\equiv\tilde{\rho}_{\rm mi }\,a^{-3}, 
\end{eqnarray}
where $\tilde{\rho}_{mi }>0$ is an integration constant. It will be shown that, even in the general case where the fractional equations are solved, this relation holds throughout all cosmological epochs.

The analysis of the solutions, as well as the fundamental role of the fractional framework in generating these regimes, is deferred to 
Subsection \ref{Analysis} and Sections \ref{unif-pot} and \ref{Concl}.

\subsection{Radiation dominated regime}
\label{Radiation}
We begin with the radiation-dominated epoch as the simplest nontrivial background, which allows us to illustrate the fractional structure of the solutions in a transparent way.

To describe a radiation-dominated epoch, we assume an effective equation of state  $w_{_{\rm eff}}=1/3$, in direct analogy to relativistic cosmology.
Assuming  $\mathcal{K}=0$, the fractional equations \eqref{eff-frd} 
and \eqref{w-gen} admit the following exact background solutions:
\begin{eqnarray}\label{Rad-a}
a_{r}(t)&=&A_r \,(t-t_r) ^{\frac{1}{2}}, \\\nonumber\\
\label{Rad-rho}
\rho^{(rad)}_{_{\rm eff}}&=&\rho_{ri} \left(\frac{a}{a_i}\right)^{-4}\equiv\tilde{\rho}_{ri }\, a^{-4}, 
\end{eqnarray}
where $\rho_{ri}=\rho_{ri}(a_i)$ is an integration constant and the constant $A_r$ is defined by
\begin{eqnarray}\label{a-r}
 A_r\equiv \, \left(\frac{32\, \pi \,G }{3} \tilde{\rho}_{ri }\right)^{\frac{1}{4}}.
\end{eqnarray}
 Moreover, $t_r$ is an integration 
constant such that we set $a(t_r)=0$. Without loss of generality, we may shift the origin of time and set $t_r=0$. 

Using \eqref{Rad-rho}, and substituting $p_{_{\rm eff}}=1/3\rho_{_{\rm eff}}$ into the equation \eqref{p-eff}, we obtain the fractional energy contribution 
\begin{eqnarray}\label{E-alph-rad}
E^{(rad)}_{\alpha}(a)=-\frac{A_r^4}{8 a^2}+\frac{C^{(rad)}_{\alpha}}{a},
\end{eqnarray}
where $C^{(rad)}_{\alpha}$ is an integration constant and vanishes when $\alpha=1$. Moreover, substituting $E^{(rad)}_{\alpha}(a)$ from \eqref{E-alph-rad} and $\rho^{(rad)}_{_{\rm eff}}$ from \eqref{Rad-rho} into the relation \eqref{rho-eff}, we obtain 
\begin{eqnarray}\label{rho-m-rad}
\rho^{\rm (rad)}_{\rm m}=\left(\frac{3\,C^{\rm (rad)}_{\alpha}}{4 \pi \,G}\right)\frac{1}{a^3}. 
\end{eqnarray}

Using equations \eqref{cos-T-def}, \eqref {cos-E-def}, \eqref{Rad-a}, \eqref{Rad-rho} and \eqref{rho-m-rad}, one can easily obtain $E$, and show that $\mathcal{K}=0$, as expected. Moreover, substituting $a_r(t)$ from \eqref{Rad-a} into equation \eqref{cos-T-def}, we obtain
\begin{eqnarray}\label{T-alpha-rad}
T^{(rad)}_{\alpha}=-\frac{(\alpha-1)C^{(rad)}_{\alpha}}{a^2}+T^{(rad)}_{\alpha *},
\end{eqnarray}
where $T^{(rad)}_{\alpha *}=T^{(rad)}_{\alpha}(a^*)$ is an integration constant such that it vanishes when $\alpha=1$.

Finally, using equations \eqref{E-alph-rad} and \eqref{T-alpha-rad}, we obtain the fractional potential as
\begin{eqnarray}\label{Phi-alpha-rad}
\Phi^{(rad)}_{\alpha}=\frac{(\alpha-\frac{3}{2})\,A_r^4}{4\, a^2}+\frac{C^{(rad)}_{\alpha}}{a}-T^{(rad)}_{\alpha *}.
\end{eqnarray}
In summary, the radiation-dominated regime provides a clear illustration of the structure of fractional NC. Although the background evolution exactly reproduces the relativistic radiation era, the underlying energy decomposition reveals a non-trivial fractional sector. 
A detailed discussion of how this regime emerges within the fractional Newtonian dynamics is presented in Subsection \ref{Analysis} and Section \ref{unif-pot}.


\subsection{Matter dominated regime}
\label{Matter-dominated}
We now consider the matter-dominated regime and derive the corresponding background quantities within the fractional NC.

Assuming a pressure-less effective fluid, $p_{_{\rm eff}}=0$, and restricting to the case $\mathcal{K}=0$, the equations \eqref{eff-frd} and \eqref{w-gen} admit the following exact solutions
\begin{eqnarray}\label{mat-a-rho}
a(t)=A_m \,(t-t_m) ^{\frac{2}{3}}, \hspace{2mm}
\rho^{(m)}_{_{\rm eff}}=\rho_i \left(\frac{a}{a_i}\right)^{-3}\equiv\tilde{\rho_i}\, a^{-3},
\end{eqnarray}
where $\rho_i=\rho(a_i)$ is an integration constant 
and for the sake of simplicity we set $a(t_m)=0$. One can also shift the origin of the time and assume $t_m=0$. Moreover, in equation \eqref{mat-a-rho}, $A_m$ is defined for convenience in terms of the initial conditions and the gravitational constant:
\begin{eqnarray}\label{A-m}
 A_m\equiv \, \Big({6 \pi \,G } \,\tilde{\rho_i}\Big)^{\frac{1}{3}}.
\end{eqnarray}

Following the same procedure as in the previous subsection, substituting $a(t)$ from \\\eqref{mat-a-rho} into equations \eqref{cos-T-def}, \eqref{cos-E-def} and \eqref{cons-eq-rho-m}, one can easily obtain the other quantities of the model as functions of the scale factor:
\begin{eqnarray}\label{E-alph-mat}
E^{(m)}_{\alpha}=\frac{C^{(m)}_{\alpha}}{a},
\end{eqnarray}
\begin{eqnarray}\label{rho-m-mat}
\rho_{\rm m}(a)=\left(\tilde{\rho_i}+\frac{3\,C^{(m)}_{\alpha}}{4 \pi \,G}\right)\frac{1}{a^3},
\end{eqnarray}
\begin{eqnarray}\label{T-alpha-mat}
T^{(m)}_{\alpha}=-\left[\frac{2}{3}(\alpha-1)A_m^3\right]\frac{1}{a}+T^{(m)}_{\alpha *},
\end{eqnarray}
\begin{eqnarray}\label{Phi-alpha-mat}
\Phi^{(m)}_{\alpha}=\left[\frac{2}{3}(\alpha-1)A_m^3+C^{(m)}_{\alpha}\right]\frac{1}{a}-T^{(m)}_{\alpha *},
\end{eqnarray}
where $C^{(m)}_{\alpha}$ and $T^{(m)}_{\alpha *}=T^{(m)}_{\alpha }(a*)$ are two integration constants, which depend on the fractional parameter $\alpha$ and vanish in the limit $\alpha=1$. 
Moreover, using the solutions associated with this case and the definitions \eqref{cos-E-def} and \eqref{K-def}, one can easily show $\mathcal{K}=0$, as expected.

We should note that the solution \eqref{rho-m-mat} is derived by solving the set of equations of our fractional model and, as expected, it is consistent with the result \eqref{dens-m}. Therefore, using equations \eqref{mat-a-rho} and \eqref{rho-m-mat}, a relation between the integration constants can be established:
\begin{eqnarray}\label{const-m}
\tilde{\rho}_i=\tilde{\rho}_{mi }-\frac{3\,C^{(m)}_{\alpha}}{4 \pi \,G}.
\end{eqnarray}
As can be seen, in the limit $\alpha=1$, we get $\tilde{\rho}_i=\tilde{\rho}_{mi }$, and consequently the standard Newtonian solution is recovered. 

The properties of these solutions and the role of fractional contributions will be analyzed in Subsection \ref{Analysis} and Section \ref{unif-pot}.

\subsection{Late time accelerating regime}
\label{L-dominated}

We finally consider a late-time accelerating regime \cite{rasouli2020late} and derive the corresponding background solutions within the fractional NC.

We assume a constant effective pressure driving the accelerated expansion:
\begin{eqnarray}\label{L-rho}
p_{_{\rm eff}}=-\rho_{_{\rm eff}}=-\rho_{\Lambda}=\rm{constant}<0, 
\end{eqnarray}
Substituting the effective energy density \eqref{L-rho} into the fractional 
Friedmann equation \eqref{eff-frd}, we obtain
\begin{eqnarray}\label{L-a}
a(t)=A_{de} \,e^{H_{de}t}, \hspace{10mm}
H_{de}\equiv \sqrt{\frac{8 \pi G}{3}\rho_{\Lambda}},
\end{eqnarray}
where we assumed $\mathcal{K}=0$ and $A_{de}$ is an integration constant.
It is straightforward to show that the fractional energy contribution and the ordinary matter density associated with the accelerating regime can be expressed as a function of the scale factor as
\begin{eqnarray}\label{E-alph-de}
E^{(de)}_{\alpha}=-\left(\frac{4 \pi G\,\rho_{\Lambda}}{3}\right)a^2+\frac{C^{(de)}_{\alpha}}{a},
\end{eqnarray}
\begin{eqnarray}\label{rho-m-de}
\rho^{\rm (de)}_{\rm m}=\frac{3\,C^{\rm (de)}_{\alpha}}{4 \pi \,G}\frac{1}{a^3},
\end{eqnarray}
where we used equations \eqref{cos-T-def}, \eqref{cos-E-def},  \eqref{rho-eff}, \eqref{cons-eq-rho-m} and \eqref{L-a}.
To obtain the quantity $T_\alpha(t)$ associated with this case, we substitute the scale factor from \eqref{L-a} into \eqref{cos-T-def}, which yields
\begin{eqnarray}\label{T-alpha-de-t}
T^{(de)}_{\alpha}(t)\simeq-\left[\frac{1}{2}(\alpha-1) H_{de}\,a_i^2\right]\left(\frac{e^{2 H_{de}\,t}}{t}\right)+B_{\alpha },
\end{eqnarray}
where, we assumed $t\gg1$, and $B_{\alpha }$ is an integration constant that should be set equal to zero for $\alpha=1$. Moreover, to obtain the relation \eqref{T-alpha-de-t}, we employed the following procedure: 
\begin{eqnarray}\label{E-i}
\int^t_{t_*}\frac{e^{2 H_{de}\,t}}{t}=E_i\left(2 H_{de}\,t\right),
\end{eqnarray}
where the exponential integral is evaluated between a reference time $t_*$ and $t\gg1$, and $E_i(x)$ is a special function:
\begin{eqnarray}\label{E-i}
E_i\left(2 H_{de}\,t\right)\simeq e^{2 H_{de}\,t} \left[\frac{1}{2 H_{de}\,t}+\left(\frac{1}{2 H_{de}\,t}\right)^2\right].
\end{eqnarray}

Using equation \eqref{L-a}, one can easily write $T^{(de)}_{\alpha}$ in terms of the scale factor:

\begin{eqnarray}\label{T-alpha-de}
T^{(de)}_{\alpha}=\left[\frac{1}{2}(\alpha-1)(a_i\,H_{de})^2\right]\frac{(\frac{a}{a_i})^2}{\ln(\frac{a}{a_i})}+B_{\alpha }.
\end{eqnarray}
Finally, using relations \eqref{E-alph-de} and \eqref{T-alpha-de}, one can easily obtain the fractional potential associated with this case:
\begin{eqnarray}\nonumber
\Phi^{(de)}_{\alpha}(a)&=&-\left(\frac{4 \pi G\,\rho_{\Lambda}}{3}\right)a^2+\frac{C^{(de)}_{\alpha}}{a}\\\nonumber\\
\label{Phi-alpha-de}
&-&\left[\frac{1}{2}(\alpha-1)(a_i\,H_{de})^2\right]\frac{(\frac{a}{a_i})^2}{\ln(\frac{a}{a_i})}
-B_{\alpha }.
\end{eqnarray}

The analysis of these solutions and their interpretation are presented in  Subsection \ref{Analysis} and Section \ref{unif-pot}.


\subsection{Analysis of the properties of the solutions}
\label{Analysis}

We now analyze the main physical features of the cosmological solutions derived in the previous subsections.

\begin{itemize}

\item 
Let us first present our strategy for deriving the background solutions.
For each cosmological phase, we assigned an effective equation of state inspired by the corresponding relativistic models.
The effective fluid description was adopted at the background level. The first step consisted in solving the continuity equation for the effective fluid, which uniquely determines the effective energy density as a function of the scale factor.
Then, the resulting expression for $\rho_{\rm eff}(a)$ was substituted into the fractional Friedmann equation \eqref{eff-frd}, where we assumed $\mathcal{K}=0$.
From this equation, the scale factor $a(t)$ was obtained in a closed analytic form. Finally, using these results, we obtained the other relevant quantities of our model.

An important result is that the exact solutions derived in this way satisfy all defining equations of the fractional model simultaneously. 
\item In particular, the fractional Friedmann equation \eqref{eff-frd}, the acceleration equation \eqref{acc-eff-eq}, and the continuity equations \eqref{cons-eq-cos}-\eqref{cons-eq-rho-m} form a closed, self-consistent dynamical system.
No internal inconsistency arises between all the definitions associated with the quantities of the fractional model.
This confirms the internal coherence of our fractional framework at the background level. In summary, at the effective level, we obtained the corresponding relativistic cosmological models with vanishing $\mathcal{K}$. No ad hoc tuning or artificial suppression of the matter sector is required. Moreover, the proposed framework admits a smooth Newtonian limit.
Concretely, as $\alpha = 1$, all fractional contributions automatically disappear and the model continuously reduces to standard NC.

\item
Within the framework of fractional NC, the corresponding solutions of each phase adapt a strategy different from the one described above: The phases are obtained by appropriately choosing the fractional gravitational potentials $\Phi^{(rad)}_{\alpha}$, $\Phi^{(m)}_{\alpha}$, and $\Phi^{(de)}_{\alpha}$, respectively, either in the presence or in the absence of the standard Newtonian potential. Importantly, all these solutions emerge dynamically from the same fractional formulation, without the need to introduce ad hoc modifications to the equations of motion. More generally, one may introduce an effective gravitational potential of the form
\begin{equation}
\Phi_{\rm eff}=\Phi_{\rm eff}\!\left(\Phi_{\rm N},\alpha,a\right),
\end{equation}
subject to the physically well-motivated requirement that it continuously reduces to the Newtonian potential in the limit $\alpha=1$. Furthermore, by imposing suitable phase-dependent conditions on $\Phi_{\rm eff}$, the effective potential naturally reduces to the fractional potentials $\Phi^{(rad)}_{\alpha}$, $\Phi^{(m)}_{\alpha}$, and $\Phi^{(de)}_{\alpha}$ in the radiation-, matter-, and acceleration-dominated eras, respectively. This construction provides a unified and self-consistent description of evolution of the universe within the fractional NC, highlighting the unifying character of the model. Such a strategy will be used in the following section.

\item
In standard NC, background solutions corresponding to radiation domination or late-time accelerated expansion cannot be obtained unless additional terms are introduced by hand into the equations of motion \cite{Gouba:2021bjs,arcuri1994growth,vieira2014lagrangian}. In sharp contrast, the fractional NC developed here naturally accommodates radiation-, matter-, and acceleration-dominated eras within a unified dynamical framework. 

\item 
At late times, our fractional model reproduces a de Sitter-like
accelerated regime with 
$a(t)\sim e^{H_\alpha t}$, $\rho_{\rm eff}\to \mathrm{const.}$ and 
$w_{\rm eff}\to -1$, thus mimicking the background dynamics of a 
$\Lambda$CDM universe for a spatially flat FLRW metric. The crucial difference is that, in our
framework, the accelerated expansion does not originate from an explicit 
cosmological constant term in Einstein's equations or from a phenomenological
dark-energy fluid, but rather from fractional modifications of the mechanical
energy, encoded in the kinetic and potential sectors associated with the 
parameter $\alpha$. This 
identifies $\alpha$ as a structural deformation parameter interpolating between 
purely Newtonian dynamics and relativistic-like accelerated expansion, and 
opens the possibility of constraining $\alpha$ directly from late-time
observations. Furthermore, the fractional effective density and pressure satisfy 
their own continuity equations, while the ordinary matter component obeys the 
standard Newtonian continuity law. Hence, the fractional and matter sectors are 
naturally decoupled at the background level, providing a mechanical 
interpretation of a dark-energy-like component emerging from fractional dynamics rather than being postulated at the level of spacetime geometry.

\end{itemize}

\section{Constructing a unified interpolating effective potential}
\label{unif-pot}

In the previous section, we have shown that our fractional 
framework is able to reproduce the exact background 
solutions corresponding to the $\Lambda$CDM model. 
The method can be summarized as follows. By considering a single effective potential of the form
\begin{equation}
\Phi_{\rm eff}(\Phi_N,a,\alpha)
=\Phi_N(a)+\sum_{i=1}^3\Phi_{\alpha i}(a),
\end{equation}
where the fractional potentials $\Phi_{\alpha i}(a)$ are given by the corresponding equations \eqref{Phi-alpha-rad}, \eqref{Phi-alpha-mat} and \eqref{Phi-alpha-de} dominating according to the cosmological epoch under consideration. This objective, together with the inclusion of the friction-like term generated by the time-dependent kernel and the explicit presence of ordinary matter in all cosmological regimes (in particular during the radiation era and the current accelerated phase, despite their subdominant contributions in these epochs), naturally leads to a rather involved structure for the potential. Nevertheless, based on the heuristic forms of the fractional potentials \eqref{Phi-alpha-rad}, \eqref{Phi-alpha-mat} and \eqref{Phi-alpha-de}, simpler and more minimal forms of the effective potential can also be considered, which will be discussed below.
We will show that this strategy not only facilitates a simpler reconstruction of the dynamics in each regime, but also provides a transparent mechanism to determine theoretical constraints on the fractional parameter $\alpha$ and, moreover, to extract observational constraints on its allowed range.

 In this section, our aim is to propose an effective
interpolating potential $\Phi_{\rm eff}(a;\alpha)$ that
continuously connects the radiation, matter, and late-time accelerating
eras without introducing ad hoc phase transitions. 
The key physical requirement is that $\Phi_{\alpha}$ must activate or
deactivate depending on the cosmological regime. 
Motivated by these criteria, we introduced two smooth switching
functions (window functions) $W_{\rm rad}(a)$ and $W_{\Lambda}(a)$ that govern how fractional terms are turned on and off across cosmic time.

We consider a new interpolated fractional potential as
\begin{equation}\label{par-phi-alpha}
\Phi_{\alpha}(a;\alpha)=R(\alpha)\,a^{-2}\,W_{\rm rad}(a)+L(\alpha)\,a^{2}\,W_{\Lambda}(a),
\end{equation}
where the amplitudes $R(\alpha)$ and $L(\alpha)$ are determined by
observational constraints (radiation era and late-time fits,
respectively), and
\begin{equation}\label{switch}
W_{\rm rad}(a)
\equiv \frac{1}{1+\bigl(\tfrac{a}{a_{\rm eq}}\bigr)^{n_r}},
\qquad
W_{\Lambda}(a)
\equiv \frac{1}{1+\bigl(\tfrac{a_{\Lambda}}{a}\bigr)^{n_\Lambda}},
\end{equation}
with $n_r,n_\Lambda>0$ controlling the sharpness of the transitions.
The scale factor is taken dimensionless with $a_0=1$ today, and
$a_{\rm eq}$, $a_{\Lambda}$ denote the equality and acceleration
transition scales, respectively. Moreover, we will assume $\mathcal{K}=0$.

We should stress that the window functions introduced in equation \eqref{switch} are not intended to represent independent fractional contributions. Rather, they are phenomenological smooth switching functions introduced to construct a single unified effective potential capable of interpolating continuously between the radiation-dominated, matter-dominated, and late-time accelerating regimes. The functional structure of the interpolating potential in equation \eqref{par-phi-alpha} is directly motivated by the exact fractional potentials previously derived for the individual $\Lambda$CDM cosmological phases. 
The fractional character of the framework is encoded in the effective fractional potential $\Phi_\alpha(a;\alpha)$ itself, which explicitly depends on the deformation parameter $\alpha$ and continuously vanishes in the standard Newtonian limit $\alpha=1$. In contrast, the role of the window functions $W_{\rm rad}(a)$ and $W_{\Lambda}(a)$ is only to provide a smooth activation or suppression of the corresponding fractional terms across different cosmological epochs, ensuring the correct asymptotic behavior in each regime.

It is important to note that the unified effective potential 
introduced in equation \eqref{par-phi-alpha}
should not be interpreted as a strictly unique functional form. 
Rather, it represents one of the simplest realizations within a broader class 
of effective potentials capable of interpolating between the radiation–dominated, 
matter–dominated, and accelerated regimes in the fractional Newtonian framework. 
Different potentials sharing the same asymptotic structure and scaling behavior 
are expected to lead to essentially the same background cosmological dynamics. 
In this sense, the form adopted here may be regarded as a minimal representative 
of a wider family of potentials whose leading-order expansion yields the same 
$\Lambda$CDM-like evolution derived in the present work.

\subsection{Radiation-dominated phase with the unified effective potential}
\label{rad-particular-pot}

In this era, as $a\ll a_{\rm eq}$, using the relations \eqref{switch}, we get $W_{\rm rad}\!\to\!1$ and $W_{\Lambda}\!\to\!0$. Therefore,  equations \eqref{eff-pot-cos-0} and \eqref{par-phi-alpha} yield 

\begin{eqnarray}\label{eff-pot-cos}
\Phi_{\rm eff}=\Phi_{_{N}}+ R(\alpha)\,a^{-2}.
\end{eqnarray}

During the radiation era ($a\ll a_{\rm eq}$), the scaling
$\Phi_{N}$ makes it dynamically subdominant with respect to the
fractional contribution $\Phi_{\alpha}\propto a^{-2}$. Therefore, equation \eqref{acc-eq} reduces to 
\begin{eqnarray}\label{rad-acc-eq}
 \ddot{a}(t)+(\alpha-1)\frac{\dot{a}(t)}{t}-R(\alpha)\,a^{-3}\simeq0.
\end{eqnarray}

One can easily show that equation \eqref{rad-acc-eq} admits the following exact solution:
\begin{equation}\label{rad-a-A}
a_{\rm rad}(t)
= A(\alpha)\, t^{1/2},
\qquad
A_{\rm rad}(\alpha)
= \left[\frac{4\,R(\alpha)}{\alpha - \tfrac{3}{2}}\right]^{1/4}.
\end{equation}
 In our fractional cosmological framework, the standard Newtonian regime is
recovered in the limit $\alpha = 1$. Consequently, the theoretical domain of
$\alpha$ in equation \eqref{rad-a-A} must be restricted to $\alpha < \tfrac{3}{2}$. Within this interval, the requirement that the coefficient $A_{\rm rad}(\alpha)$ remains real and positive imposes the additional condition $R(\alpha) < 0$.

Substituting $a_{\rm rad}(t)$ into \eqref{eff-frd}, we obtain

\begin{eqnarray}\label{rho-rad-part}
 \rho_{_{\rm eff}}\simeq \rho_{\alpha}=
 \frac{3}{32\, \pi\, G} \left(\frac{A_{\rm rad}}{a_{\rm rad}}\right)^4,
\end{eqnarray}
where we assumed $\mathcal{K}=0$. 

Using equations \eqref{cos-T-def}, \eqref{rho-eff}, \eqref{p-eff}, \eqref{rad-a-A} and \eqref{rho-rad-part}, we can easily obtain the other quantities associated with the radiation-dominated phase as
\begin{eqnarray}\label{E-alpha-part-pot}
 E_{\alpha}&=&
 - \frac{A^4_{\rm rad}}{8\,a^2_{\rm rad}},\qquad T_{\alpha}=
 \frac{A^4_{\rm rad}(1-\alpha)}{4\,a^2_{\rm rad}},\\\nonumber\\
 \label{p-rad-part}
  p_{_{\rm eff}}&= &p_{\alpha}=
 \frac{1}{32\, \pi\, G} \left(\frac{A_{\rm rad}}{a_{\rm rad}}\right)^4.
\end{eqnarray}
From \eqref{rho-rad-part} and \eqref{p-rad-part}, we obtain $w_{\rm eff}=1/3$, and one can easily check that all the equations and definitions of the model remain consistent with these exact solutions. 

As follows from equation~\eqref{rad-a-A}, it is seen that the fractional corrections do \emph{not} modify the
time-exponent, and the standard scaling $a_{\rm rad}\propto t^{1/2}$ is preserved 
in agreement with $\Lambda$CDM model.
 The fractional sector influences only
the normalization $A(\alpha)$, which can be constrained observationally.


\subsection{Matter-dominated phase with the unified effective potential}
\label{mat-particular-pot}

At first sight, the matter-dominated solution obtained in subsection \ref{Matter-dominated} may appear trivial, since similar power-law behaviors can also be recovered within standard NC. However, in what follows, we present a clear interpretation. Our solutions not only reproduce the standard Newtonian limit, but are also fully consistent with the corresponding background evolution of the $\Lambda$CDM model. Moreover, although the exponents of time and scale factor in the matter-dominated phase coincide with those of the standard models, the coefficients appearing in the scale factor, energy density, effective pressure, and other related physical quantities explicitly depend on the fractional parameter $\alpha$. This feature opens the possibility of non-trivial phenomenological consequences when the model is compared with generalized relativistic frameworks and cosmological observations.

In the matter-dominated regime, we can work with an effective potential as $\Phi_{\rm eff}=\Phi_{_{N}}$.
In what follows, we show that the solutions associated with the matter-dominated epoch are still recovered even in the absence of the fractional potential $\Phi_{\alpha}$, as it effectively behaves like the Newtonian potential; see Subsection \ref{Matter-dominated}. 
Equation~\eqref{Phi-alpha-mat} then immediately implies 
\begin{eqnarray}\label{C-alpha-mat-par}
C^{(m)}_{\alpha}=-\frac{2}{3}(\alpha-1)A_m^3,
\end{eqnarray}
where, without loss of generality, we assumed  $T^{(m)}_{\alpha *}=0$ . 
Therefore, in this particular case, the solutions \eqref{mat-a-rho} take the following form
\begin{eqnarray}\label{mat-a-special}
a(t)&=&\left(\frac{6 \pi \,G \tilde{\rho}_{mi } }{4-3 \alpha}\right)^{\frac{1}{3}} \,t ^{\frac{2}{3}},\\\nonumber\\
\label{mat-rho-special}
\rho^{(m)}_{_{\rm eff}}&=&\frac{\tilde{\rho}_{mi }}{4-3 \alpha} \left(\frac{1}{a}\right)^{3}, \hspace{10mm} p^{(m)}_{_{\rm eff}}=0,
\end{eqnarray}
where $\tilde{\rho}_{mi }>0$ is given by \eqref{dens-m}. 
Moreover, for this particular case,  equations \eqref{E-alph-mat} and \eqref{T-alpha-mat} yield:
\begin{eqnarray}\label{E-alph-mat-par}
E^{(m)}_{\alpha}=T^{(m)}_{\alpha}=-\left[\frac{4 \pi G (\alpha-1)}{4-3 \alpha}\right]\frac{1}{a}.
\end{eqnarray}

It is seen that the exponents of the time and scale-factor appearing in solutions \eqref{mat-a-special} and \eqref{mat-rho-special} coincide exactly with those of the standard Newtonian and $\Lambda$CDM cosmological models. 
 However, the corresponding coefficients become explicit functions of the fractional parameter $\alpha$ and reduce to the standard Newtonian expressions in the limit $\alpha=1$. 

We emphasize that the emergence of the fractional parameter $\alpha$ in the equations and solutions of the model is rooted in the presence of the time-dependent kernel and the fractional potential $\Phi_\alpha$. 
In the special case where $\Phi_\alpha = 0$, the influence of the time kernel remains and is encoded entirely in the form of time-memory effects.
In the general case $\alpha \neq 1$, constraints on the fractional parameter may be obtained by comparing these background solutions with their counterparts in the $\Lambda$CDM model and with cosmological observations.
This demonstrates that, within our fractional framework, genuinely fractional effects already manifest themselves at the background level. 
In addition, the coefficient in equation \eqref{mat-a-special} must remain positive, which restricts the allowed range of the fractional parameter to $\alpha<4/3$.


\subsection{Late-time fractional cosmology with the unified effective potential}

In this regime,  using relations \eqref{switch}, we get $W_{\rm rad}\!\to\!0$ and $W_{\Lambda}\!\to\!1$. Therefore,  equations \eqref{eff-pot-cos} and \eqref{par-phi-alpha} yield 

\begin{eqnarray}  \label{Phi-alpha-def}
\Phi_{\rm eff}=\Phi_{_{N}}+ L(\alpha)\,a^{2}.
\end{eqnarray}

At sufficiently large $a$, the Newtonian potential can be 
neglected compared to $ L(\alpha)\,a^{2}$. Nevertheless, despite the fact that the term emerges from the time-dependent kernel is strongly suppressed at late
times, we keep it in our analysis so as to capture its residual impact and to
evaluate potential sub-leading corrections to the asymptotic dynamics. Moreover, let us consider the case with $\mathcal{K}=0$.

Ultimately, under the assumptions stated above, the equation \eqref{acc-eq} reduces to

\begin{equation}\label{acc-late}
  \ddot{a}(t) +\frac{\alpha - 1}{t}\,\dot{a}(t)-\kappa^2\, a(t)= 0,\qquad \kappa^2(\alpha) \equiv -2 L(\alpha)>0.
\end{equation}

One can show that the general solution of equation \eqref{acc-late} can be written as
a linear combination of modified Bessel functions $I_{\nu}$ and $K_\nu$ of order $\nu$, where 
$\nu \equiv \left|1 - \frac{\alpha}{2}\right|$.
However, in the limit of sufficiently large times, the asymptotic behavior of these  Bessel functions
indicates that the following expression provides an accurate approximation for
describing the current accelerated universe:

\begin{eqnarray}
  a(t)  &\simeq & A_{\rm L}\, t^{\frac{1-\alpha}{2}} e^{\kappa t}\left[1+\frac{c_2}{t}+\mathcal{O}\!\left(\frac{1}{t^2}\right)\right],
  \label{a-late-time-part}
   \end{eqnarray}
    where  $A_{\rm L}\equiv c_1\,\sqrt{\frac{1}{2\pi \kappa}}$,
  $\kappa t\gg 1$, and $c_1$ and $c_2$ are integration constants.
Equation~\eqref{a-late-time-part} indicates that the model admits a de Sitter-like attractor, modulated by a mild power-law prefactor that depends on the fractional parameter $\alpha$.

Using  the asymptotic solution~\eqref{a-late-time-part} and the fractional equations \eqref{cos-T-def}-\eqref{p-eff} given in Section \ref{Frac-Cosmology}, it is straightforward to
show that the other quantities in the late-time regime take the following form:
\begin{equation} \label{H-late-time-part}
  H(t) \simeq\kappa + \frac{1-\alpha}{2t} + \mathcal{O}\!\left(\frac{1}{t^2}\right),
  \qquad
 \kappa t\gg 1,
 \end{equation}
 \begin{equation}\label{rho-alpha-late-part}
  \rho_{\rm eff}(t)\simeq \rho_\alpha(t)
  = \frac{3}{8\pi G}
  \left[
    \kappa^2
    + \frac{\kappa(1-\alpha)}{t}
    + \mathcal{O}\!\left(\frac{1}{t^2}\right)
  \right],
\end{equation}
\begin{equation}\label{p-alpha-late-part}
  p_\alpha(t)
  \simeq -\frac{1}{8\pi G}
  \left[
    3\kappa^2
    + \frac{3\kappa(1-\alpha)}{t}
    + \mathcal{O}\!\left(\frac{1}{t^2}\right)
  \right],
  \end{equation}

\begin{equation} \label{Talpha-late-part}
  T_\alpha(t)
  \simeq \frac{(\alpha-1) \kappa \, a^2(t)}{2\,t}
 \end{equation}
 \begin{equation}  \label{E-alpha-late-part}
  E_\alpha(t)
    \simeq -\frac{1}{2}a^2(t)
        \left[
          \kappa^2 + \frac{\kappa(1-\alpha)}{t}
        \right].
\end{equation}

In the remainder of this subsection, our goal is to investigate how the physical quantities associated with the accelerating regime of the fractional model can be compared with the corresponding behavior of the $\Lambda$CDM model, as well as with recent observational constraints. More specifically, we analyze how the effective equation of state parameter, the effective cosmological constant, and the Hubble rate arising in the present framework are related to their corresponding quantities in standard relativistic cosmological models and how they may be physically interpreted.

We estimate a rough observational constraint on $\alpha$ from the 
effective equation of state at the present epoch. 
Using~\eqref{eff-frd} and~\eqref{acc-eff-eq}, the effective equation of state parameter can be written as
\begin{equation}\label{w-H}
  w_{\rm eff}(t)=\frac{p_{\rm eff}}{\rho_{\rm eff}}
  = -1 - \frac{2}{3}\frac{\dot{H}}{H^2}.
\end{equation}
Substituting $H(t)$ from~\eqref{H-late-time-part} into \eqref{w-H}, we obtain
\begin{equation}\label{w-t-late}
  w_{\rm eff}(t)
 \simeq -1 + \frac{1-\alpha}{3\kappa^2 t^2}
      + \mathcal{O}\!\left(\frac{1}{t^3}\right).
\end{equation}
Evaluating this expression at the present time $t=t_0$, and assuming that the asymptotic rate $\kappa$ is of the order of the current Hubble parameter $H_0$, we obtain
\begin{equation}
  w_0 \equiv w_{\rm eff}(t_0)
  \simeq -1 + \frac{1-\alpha}{3H_0^2 t_0^2}.
\end{equation}
Planck 2018 and late-time probes constrain the constant EoS parameter $w_0$ to be very close to $-1$, typically $w_0 \simeq -1.028 \pm 0.031$ (at the $1\sigma$ level) \cite{Planck:2018vyg} (ee also \cite{nagpal2025late,chaudhary2025lambdacdm,Chaudhary:2025vzy} for recent analyses of evolving dark energy scenarios and possible deviations from the standard $\Lambda$CDM cosmology), so that at the $2\sigma$ level, one can conservatively require
\begin{equation}\label{obs-B}
  |w_0 + 1| \lesssim 0.06.
\end{equation}
Using the fact that in a $\Lambda$CDM-like Universe one has $H_0 t_0 \sim \mathcal{O}(1)$ (numerically $H_0 t_0 \simeq 0.95$), equation \eqref{w-t-late} yields
\begin{equation}
  |w_0 + 1|
  \simeq \frac{|1-\alpha|}{3H_0^2 t_0^2}
  \sim \frac{|1-\alpha|}{3},
\end{equation}
so that the observational bound \eqref{obs-B} translates into $  |1-\alpha|
  \lesssim 0.2$, which yields 
  \begin{equation}\label{alpha-obs}
  0.8 \;\lesssim\; \alpha \;\lesssim\; 1.2.
\end{equation}

Regarding the late-time de Sitter-like regime of the model, inequality \eqref{alpha-obs} indicates that the fractional parameter must be close to unity to be consistent with current bounds on $w_0$. It should be noted that this allowed range for $\alpha$ is fully consistent with the constraints obtained in the previous subsections. 

For large values of $t$, equation \eqref{rho-alpha-late-part} can be written as 
\begin{equation}\label{rho-Lambda-eff}
  \rho_{_{\rm eff}}
 \simeq\frac{3\kappa^2(\alpha)}{8\pi G}\equiv\frac{\Lambda_{\rm eff}(\alpha)}{8\pi G}={\rm constant},
  \end{equation}
where $\Lambda_{\rm eff}(\alpha)\gtrsim0$ acts as an effective cosmological constant, which according to \eqref{acc-late}, depends on $\alpha$: ${\Lambda_{\rm eff}}={\Lambda_{\rm eff}}(\alpha)$.
Therefore, the late-time fractional regime is asymptotically de Sitter, with small $\alpha$-dependent deviations that vanish for $t\gg1$.

Therefore, equations \eqref{acc-late} and ~\eqref{rho-Lambda-eff} yield
\begin{equation} \label{L-lambdaeff}
 L(\alpha)
  = -\frac{\kappa^2}{2}
  = -\frac{\Lambda_{\rm eff}(\alpha)}{6}.
 \end{equation}
Since the entire acceleration is sourced by the fractional sector, it is natural to regard $\Lambda_{\rm eff}$ as an emergent quantity induced by the fractional contribution (we should note that within the standard Newtonian cosmology, one cannot describe the late time acceleration). 

Since the fractional parameter $\alpha$ is dimensionless, the effective cosmological constant can naturally be written as the product of a cosmological scale carrying the appropriate dimensions and a dimensionless function of $\alpha$. Therefore, one may generally write
\begin{equation}\label{Lam-lam}
\Lambda_{\rm eff}(\alpha)\equiv|\lambda\,f(\alpha)|,
\end{equation}
where $\lambda$ denotes a constant cosmological scale with the same 
dimensions as $\Lambda_{\rm eff}$, while $f(\alpha)$ is a dimensionless function of the fractional parameter.

On the other hand, in the standard limit $\alpha\rightarrow1$, the fractional sector disappears and the conventional cosmological dynamics is recovered. Consequently, one naturally expects $f(1)=0$.
Within this effective description, the behavior of $f(\alpha)$ near the classical limit $\alpha=1$ is expected to capture the leading fractional corrections to the standard cosmological dynamics. Assuming that $f(\alpha)$ is sufficiently smooth in the neighborhood of $\alpha=1$, it can be expanded around this point as 
\begin{equation}\label{Tay-f}
f(\alpha)=f(1)+(\alpha-1)\left.\frac{df}{d\alpha}\right|_{\alpha=1}
+\frac{1}{2}(\alpha-1)^2 \left.\frac{d^2f}{d\alpha^2}\right|_{\alpha=1}+\mathcal{O}\!\left((\alpha-1)^3\right).
\end{equation}
Using the condition $f(1)=0$, the leading-order contribution becomes $f(\alpha)\simeq f'(1)(\alpha-1)$, where $f'(1)\equiv\left.{df}/{d\alpha}\right|_{\alpha=1}$.
Hence,
\begin{equation}\label{lambda-eff-f}
\Lambda_{\rm eff}\simeq  |\lambda\,f'(1)(\alpha-1)|.
\end{equation}
Since $f'(1)$ is a finite dimensionless constant, it can be absorbed into a redefinition of the cosmological scale. Defining the positive cosmological scale $\Lambda\equiv |\lambda\,f'(1)|$, equation \eqref{lambda-eff-f}, at leading order, gives
\begin{equation}\label{lambda-eff-lambda}
\Lambda_{\rm eff}\simeq |1-\alpha|\,\Lambda.
\end{equation}
This linear behavior is not restricted to a specific 
choice of $f(\alpha)$, but remains valid for a broad 
class of smooth functions near $\alpha=1$, such 
as $f(\alpha)=1-\alpha$, $f(\alpha)=\ln\alpha$,  $f(\alpha)=1-\alpha^n$, $f(\alpha)=e^{\alpha-1}-1$. 
Furthermore, as we have shown above, recent observational constraints 
indicating that $|1-\alpha|\ll1$ naturally imply a small 
effective cosmological constant in the present 
cosmological epoch, in qualitative agreement with the observed late-time accelerated expansion.

By combining equations~\eqref{Phi-alpha-def}, \eqref{L-lambdaeff} and \eqref{lambda-eff-lambda}, the form of the proposed fractional potential in this regime becomes fixed:
\begin{equation}\label{phi-alpha-lambda}
  \Phi_\alpha(a)
  = -\left(\frac{\left|1-\alpha\right|}{6}\right)\,\Lambda\,a^2,
\end{equation}
and the late-time Hubble constant in our fractional model is:
\begin{equation}\label{H_0}
  H_0\simeq\kappa(\alpha)
  = \sqrt{\frac{\Lambda_{\rm eff}(\alpha)}{3}}
  = \sqrt{\frac{\left|1-\alpha\right|}{3}\Lambda}.
\end{equation}

Equations \eqref{lambda-eff-lambda} and \eqref{H_0} indicate that the effective cosmological
constant and the present Hubble rate are both determined by the
deviation of the fractional parameter from the Newtonian limit
$\alpha = 1$. In this sense, the magnitude of the late-time cosmic
acceleration is directly governed by the small parameter $|1-\alpha|$,
showing that the accelerated expansion arises from a small fractional
deformation of Newtonian gravity.


\section{Summary and discussions}
\label{Concl}

In this work, motivated by both physical and geometric considerations, we have constructed a minimal deformation of the classical (Newtonian) action by introducing a power-law time-dependent kernel governed by a fractional parameter $\alpha$. This kernel admits a natural interpretation as a nontrivial integration measure in time, placing the formulation within the broader class of measure-based approaches to anomalous or fractal dynamics \cite{calcagni2010quantum,calcagni2010fractal}. Within this framework, we have shown that a conserved effective quantity can still be defined, whose structure incorporates a time-averaged contribution of the standard kinetic energy, providing a natural interpretation of fractional kinetic energy.

Motivated by the emergence of a fractional kinetic term and the existence of a conserved effective quantity, we further assumed that the standard gravitational potential may consistently be generalized into an effective potential, such that it is smoothly reduced to its Newtonian counterpart in the limit $\alpha = 1$. In the particular case where the standard and fractional contributions to the potential energy are separable, a fractional counterpart of the mechanical energy can be defined, whose sum with the standard contribution reproduces the conserved quantity obtained from the equations of motion. In the special limit $\alpha = 1$, the generalized action continuously reduces to its standard Newtonian counterpart, and the equations of motion reduce to their Newtonian form.

We have shown that only in the simultaneous presence of both the time kernel and the fractional potential does the resulting framework reproduce the main background dynamical features of relativistic cosmology.
In this fractional cosmology, the effective density and  pressure satisfy a continuity equation. Moreover, due to the way the fractional mechanical energy contributes to the dynamics, the induced fractional components independently satisfy a separate continuity equation. These two continuity relations automatically recover the ordinary matter continuity equation, exactly as in standard Newtonian cosmology.

In analogy to the spatially flat case in relativistic cosmology, assuming $\mathcal{K}=0$, we demonstrated that the fractional model, with a single effective potential, consistently reproduces the expansion history associated with the radiation-dominated, matter-dominated and the present accelerated phases. We have shown that the analytical form of all resulting quantities is such that all assumptions, definitions, and dynamical equations of the fractional model are satisfied in a fully self-consistent manner.

In all three mentioned regimes, the fractional parameter $\alpha$ appears explicitly in the physical observables, allowing theoretical and observational constraints to be imposed. Our analysis shows that the simultaneous reproduction of these cosmological phases requires $\alpha$ to lie close to unity, namely, within the range $ 0.8 \;\lesssim\; \alpha \;\lesssim\; 1.2$, indicating that deviations from the Newtonian limit are small but dynamically significant.

It is worth emphasizing that within the framework of standard Newtonian
gravity it is not possible to describe the present accelerated
expansion of the universe. In the conventional Newtonian cosmological
description, the expansion dynamics governed solely by the ordinary matter
content, which inevitably leads to a decelerating cosmic expansion.

In contrast, the fractional extension of Newtonian gravity proposed
in this work provides a natural mechanism capable of reproducing the
present cosmic acceleration. As shown in equations \eqref{lambda-eff-lambda} and \eqref{H_0}, the
effective cosmological constant and the present Hubble rate are both
controlled by the deviation of the fractional parameter from the
Newtonian limit, $\alpha = 1$. Therefore, even a tiny fractional
deformation of the Newtonian dynamics can generate a non-vanishing
effective cosmological constant and hence a late-time accelerated
cosmic expansion.

From a conceptual point of view, this result establishes an interesting
bridge between Newtonian cosmology and relativistic cosmological
models. In the standard relativistic framework the accelerated
expansion is usually attributed either to a fundamental cosmological
constant or to some form of dark energy component. In the present
approach, however, the accelerated expansion emerges from a modified
gravitational dynamics rather than from the introduction of an
additional energy component. In this sense, the effective cosmological
constant appearing in our model does not represent a fundamental
vacuum energy but instead arises as a manifestation of the fractional
deformation of Newtonian gravity. In this perspective, dark energy
may be interpreted as the manifestation of a small fractional
deformation of Newtonian gravitational dynamics.

An additional noteworthy feature of the present framework is that the
relation $\Lambda_{\rm eff} \sim H_0^2$
naturally arises from the cosmological dynamics. This indicates that
the energy scale associated with the effective cosmological constant
is directly linked to the present Hubble scale. Moreover, since in the
present model $\Lambda_{\rm eff} \propto |1-\alpha|$, 
the observed smallness of the cosmological constant may be naturally
interpreted as a consequence of the fractional parameter remaining
very close to the Newtonian limit. In the limit $\alpha \to 1$, the
effective cosmological constant vanishes and the standard Newtonian
cosmological behavior without late-time acceleration is recovered.
Moreover, we should note that the radiation-dominated era 
cannot also be described within standard Newtonian cosmology. In contrast, within the present fractional framework, both regimes mentioned emerge effectively from the structure of the generalized action itself, without the need for externally imposed modifications. 



Taken together, these results suggest that the proposed fractional framework constitutes a minimal yet structurally consistent extension of Newtonian gravity. With a single fundamental parameter, it achieves theoretical coherence and observational viability across a broad range of cosmological and gravitational scales in agreement with the results of general relativity.

Within the gravitational and cosmological framework proposed in the present work, several other important problems have also been investigated separately, including the early-time evolution of the universe, such as the pre-inflationary, inflationary, and post-inflationary phases~\cite{Rasouli:2026hfy}, cosmological perturbation theory, the growth equation, and large-scale structure formation~\cite{Rasouli:2026vei}, as well as weak-gravity tests, including the perihelion precession of Mercury and the gravitational deflection of light~\cite{Rasouli:2026zyq}.
However, a detailed investigation of these topics and the presentation of their corresponding results have been beyond the scope of the present paper and have been addressed in separate studies. The main purpose of the present work has been to construct the proposed fractional framework and to investigate its cosmological behavior at the level of background cosmology.
In addition, several other research directions based on the same framework are currently under investigation, and their results will be reported in future works.

We briefly summarize the main motivation, significance, and distinctive features of the present fractional framework.
In the present work, inspired by measure-based approaches and Stieltjes-type structures in fractional dynamics, we have constructed a minimal extension of the Newtonian action which, without introducing any new dynamical degrees of freedom and solely through a single constant fractional parameter, is capable of effectively reproducing the main features of standard relativistic cosmology at the level of background cosmology. Moreover, the proposed framework is fully consistent with the correspondence principle, in the sense that, in the limit $\alpha \rightarrow 1$, the action and the dynamical equations continuously reduce to their standard Newtonian counterparts.
Unlike standard Newtonian cosmology, which is mainly capable of describing the matter-dominated epoch, the present model allows for the simultaneous reconstruction of the radiation-dominated, matter-dominated, and late-time accelerating phases of the Universe within a unified and self-consistent framework, without the need to introduce phenomenological terms by hand into the dynamical equations.
Furthermore, as mentioned, in separate works, the present framework also provides interesting results not only at the level of background dynamics, but also in the description of the early-Universe evolution, cosmological perturbations, large-scale structure formation, and weak-gravity tests. An important point is that all these features are obtained only within a parameter range very close to the standard Newtonian limit. This suggests that even a very small fractional deformation of the Newtonian action may lead to nontrivial cosmological consequences while remaining compatible with observational constraints.

From an ontological perspective, while the $\Lambda$CDM paradigm relies on several independent energy components to account for the different cosmological epochs, within the fractional Newtonian framework these behaviors emerge as distinct regimes of a single underlying gravitational structure. In this sense, the parameter $\alpha$ does not represent a new physical entity, but rather quantifies a structural deformation away from the Newtonian limit, which endogenously generates the effective dynamics associated with the various cosmic eras. 



\section{acknowledgments}
The author sincerely thanks the Reviewers for their careful reading of the manuscript and for their valuable and constructive comments.
The author acknowledges the FCT grant \textbf{UID/212/2025} Centro de Matem\'{a}tica 
e Aplica\c{c}\~{o}es da Universidade da Beira Interior, plus
the COST Actions CA23130 (Bridging high and low energies in search of
quantum gravity (BridgeQG)) and CA23115 (Relativistic Quantum Information (RQI)).




\bibliographystyle{elsarticle-num}
\bibliography{FracNewRef}

\end{document}